# Interlayer shear strength of single crystalline graphite

Ze Liu · Shoumo Zhang · Jiarui Yang · Jefferson Zhe Liu · Yanlian Yang, Quanshui Zheng

Z. Liu
Department of Engineering Mechanics and Center for Nano and Micro Mechanics, Tsinghua University, Beijing 100084, China

S.-M Zhang
Department of Engineering Mechanics and Center for Nano and Micro Mechanics, Tsinghua University, Beijing 100084, China

J.-R. Yang
Department of Engineering Mechanics and Center for Nano and Micro Mechanics, Tsinghua University, Beijing 100084, China

J. Z. Liu
Department of Mechanical and Aerospace Engineering, Monash University, Clayton, VIC 3800, Australia

Y.-L. Yang
National Center for Nanoscience and Technology, Beijing 100190, China

Y. Cheng
Department of Engineering Physics and Center for Nano and Micro Mechanics, Tsinghua University, Beijing 100084, China

Q.-S. Zheng
Department of Engineering Mechanics and Center for Nano and Micro Mechanics, Tsinghua University, Beijing 100084, China
e-mail: zhengqs@tsinghua.edu.cn

**Abstract**: Reported values (0.2 MPa ~ 7.0 GPa) of the interlayer shear strength (ISS) of graphite are very dispersed. The main challenge to obtain a reliable value of ISS is the lack of precise experimental methods. Here we present a novel experimental approach to measure the ISS, and obtain the value as 0.14 GPa. Our result can serve as an important basis for understanding mechanical behavior of graphite or graphene-based materials.

## 1 Introduction

Graphitic systems are used for a wide variety of applications, ranging from lubricant materials [1] to graphite intercalation compounds [2], graphene-based composite materials [3] and graphene oxide papers [4-5]. Graphite is also widely used as a raw material to obtain graphenes [6-7]. Despite the technological and scientific importance of graphitic systems, the knowledge of their mechanical properties, especially the interlayer shear strength (ISS), say $\tau_s$, is unexpectedly poor [8-9].

The first measurement of $\tau_s$ was done more than forty years ago by Soule and Nezbeda [10]. Their static test on highly anisotropic-annealed natural graphite gave the values of $\tau_s$ in the range 0.25-0.75 MPa, with the average value of 0.48 MPa that has been widely used in the literature [4, 11-15]. In their experiment, the basal plane shear stiffness constant $C_{44}$ was measured as 0.13-1.4 GPa. The much lower values of $C_{44}$ than those reported earlier were attributed to the basal plane dislocations [16]. The follow-up measurement by Blakslee et al. [17] on compression-annealed pyrolytic graphite yielded the value $\tau_s$ of 0.9-2.5 MPa. Surprisingly, except for the above-mentioned two works, there seem to be no any other directly measured values of ISS reported, as pointed out by Popov *et al.* [18]. The main challenge to use the conventional static test to measure the interlayer shear strength is the unavailability of sufficiently large single crystals [9]. The typical sizes of single crystalline grains in graphite are in the micrometer range and thicknesses in the nanometer range [12, 19].

Very recently, Ding *et al.*[20] used a thermal excitation method in a scanning tunneling microscope (STM) for a highly oriented pyrolytic graphite (HOPG) under compressive stress. They reported an unusual shear strength of the value 7 GPa with caution "The atomistic process for tip-sample interaction in STM might be more complicated than described here. To confirm the mechanism and to calculate the error for the experimental result of shear strength, further studies are of great help." [20]. As a fact, the same method used by Snyder *et al.* [21] to measure the ISS of HOPG lead to a value of 5 MPa, which is three orders in magnitude lower.

Theoretically, graphite presents a major challenge for quantum physics computational models due to the two completely different types of inter-atomic bonding: an exceptionally strong $sp^2$ covalent intra-layer bonding and an extremely weak van der Waals (vdw) interlayer bonding. Tight-binding atomistic simulations yielded the values of $\tau_s$ as 0.434 GPa by Bonelli et al. [24] and 0.9 GPa by Guo et al. [25]. For density functional theory (DFT), it remains notoriously difficult to describe the weak van der Waals interlayer interaction [22-23]. The standard approximations used in DFT, such as the local-density approximation (LDA) and generalized gradient approximation (GGA), cannot accurately describe long-distance interactions such as that of the van der Waals force. An alternative to standard DFT is the van der Waals density functional method, which was developed to account for the long-range interaction component by using an explicit nonlocal functional of the density. However, different vdw DFT methods have lead to very scattered results on the van der Waals interactions in graphite [27].

The above brief review calls for a non-conventional method to experimentally obtain a reliable value of the ISS for single crystalline graphite. In this paper, we introduce a novel method to estimate the ISS, which is inspired by our recent observations of the self-retracting motion phenomenon of micrometer graphite flakes on graphite mesas [11] and of the microscale superlubricity phenomenon [12]. We obtain the value of $\tau_s$ about 0.14 GPa.

## 2 Experiments

The samples that we used to study the interlayer shear strength were microscale graphite/SiO$_2$ square mesas lithographically constructed, as illustrated in Fig. 1(a), with a typical edge size ($L$) ranging from 0.5 to 20 micrometers and heights ($H$) from 100 to 500 nanometers. The fabricating process of the mesas was described in details in [11], followed the method proposed in [26]. Using a probe equipped to a micromanipulator MM3A (Kleindiek) to apply a shear force on the SiO$_2$ top surface of a mesa, as schematically illustrated in Fig. 1(b)-(c) with a straight arrow indicating the shear force, we can shear off a graphite/SiO$_2$ flake. For mesas with side lengths ($L$) smaller than 10μm, we found that many of the sheared flakes after being released can spontaneously move back to their before-sheared positions (Fig. 1(d)-(e)) [12]. We also noted that if the sheared flake was rotated at certain angles, as illustrated in Fig. 1(b)-(c) with a circular arrow indicating the torque, the two surfaces locked together [12]. These "lock-ins" occur at 60° intervals, consistent with the hexagonal symmetry of the graphite lattices [12].

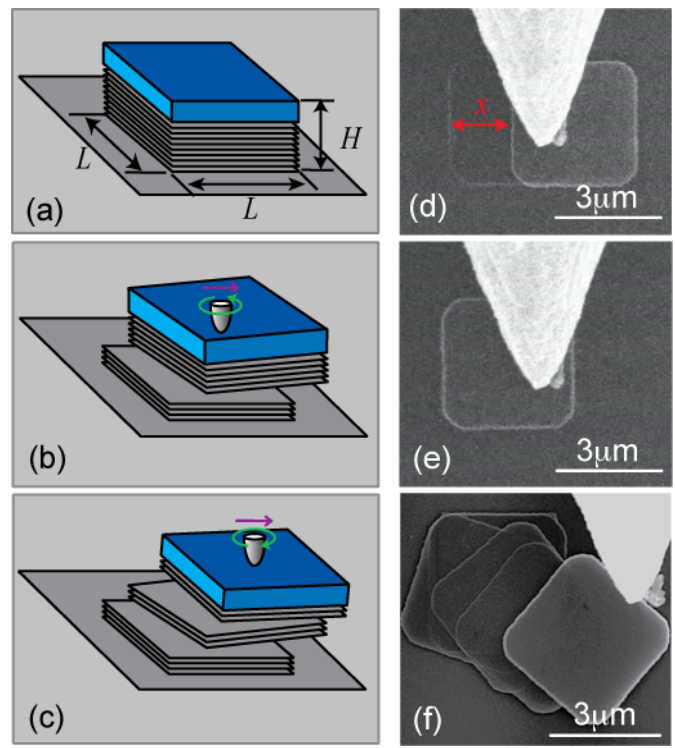

**Fig. 1** (a) Sketch of a mesa with an edge length $L$ and a height $H$. (b) Sketch of shearing out a top graphite flake and a subsequent rotation into a lock-in state. (c) Shearing the top flake again resulting in a new sliding interface. (d)-(f) SEM images: A graphite/SiO2 flake was sheared out from its platform by a microtip (d) and it can self-retract to its platform after disengaging the tip (e). (f) Multiple graphite flakes obtained by repeating the processes of (b) and (c).

Graphite is a layered material composed of single carbon layers – graphenes, in which carbon atoms are distributed in a hexagonally lattice, as illustrated in Fig. 2(a). In a single crystalline graphite, two adjacent graphenes are AB-stacked as illustrated in Fig. 2(b). The main mechanisms of the self-retraction and lock-in phenomena were revealed in [12] and summarized in following:

- Provided a contact between two self-retractable graphenes incommensurate: not only non-AB-stacked, but also have different lattice orientations as illustrated in Fig. 2(c), a negligible friction (a superlubricity state) and thus a self-retraction are observed [12].

- Provide an AB-stacking (as in a single crystal graphite), a "lock-in" state appears because of the drastically large shear resistance force per unit contact area in such a commensurate contact.

Thus, the external force required to unlock a "lock-in" state corresponds to the interlayer shear strength, $\tau_s$.

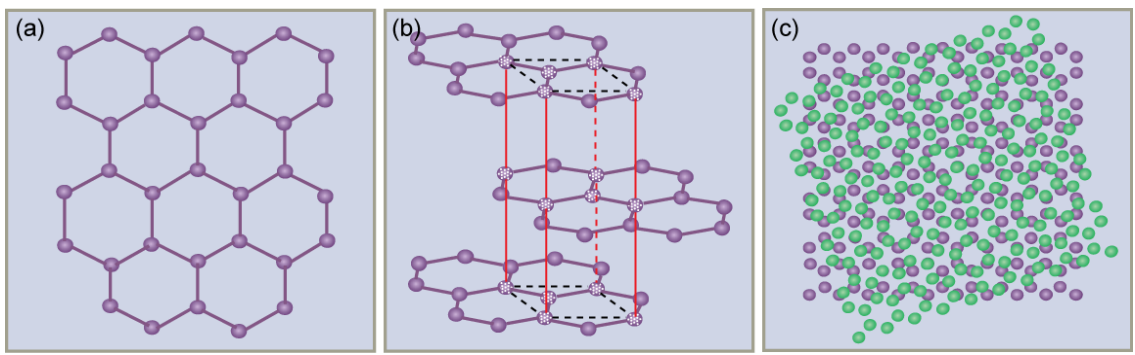

**Fig. 2** (a) Sketch of a single layer graphene with the carbon atoms in a hexagonal lattice. (b) Sketch of two graphene layers in an AB-stacking state. (c) Sketch of two graphene layers in a non-AB-stacking state generated by a relative rotation, resulting in a clear Moire pattern (different colors represent different layers). For the sake of clarity, the C-C bonds denoted by line segments in (a) are omitted.

Electron back scattered diffraction (EBSD) images (Fig. 3(a) and Fig. 3(b)) show that the HOPG sample is composed of single crystalline graphite grains with a typical size of several micrometers. Recent experiments combining focused-ion-beam/SEM and HRTEM [19] and our experiments on the self-retraction of graphite flakes [11, 12, 27] reveal a thickness of graphite single crystal grain about 5-60 nm. Based on these experimental facts, we proposed a brick-wall model for HOPG polycrystalline structure: single crystal grains (i.e., the bricks) with c-axis well aligned within few degrees but rotationally mis-oriented in the basal plane [12]. Thus adjacent graphene layers within a graphite grain are in AB-stack (commensurate) while those between different grains are most likely incommensurate. For a mesa with a few micrometers in size and about one hundred nanometers in height, there is a certain probability of existing a grain boundary or equivalently an incommensurate contact that crosses all over the mesa. We found that all self-retractable contacts are atomically smooth and incommensurate [12] (Fig. 3(c) and 3(d)). The resistant shear strength against the sliding motion determined in our experimented is very small, with an upper bound of 0.3 MPa. When continuously rotating a self-retractable flake, it will eventually been locked in. Then, when we try to shear the flake again, it is often observed that a new flake is sheared out at another incommensurate contact. Repeating the above shearing (to open an incommensurate contact) and rotating (to lock-in) process (Fig. 1(b)-(c)), we can ensure that all the contacts in the graphite mesa are AB-stacking, as illustrated in Fig.1(f), which is crucial for measuring the ISS of graphite.

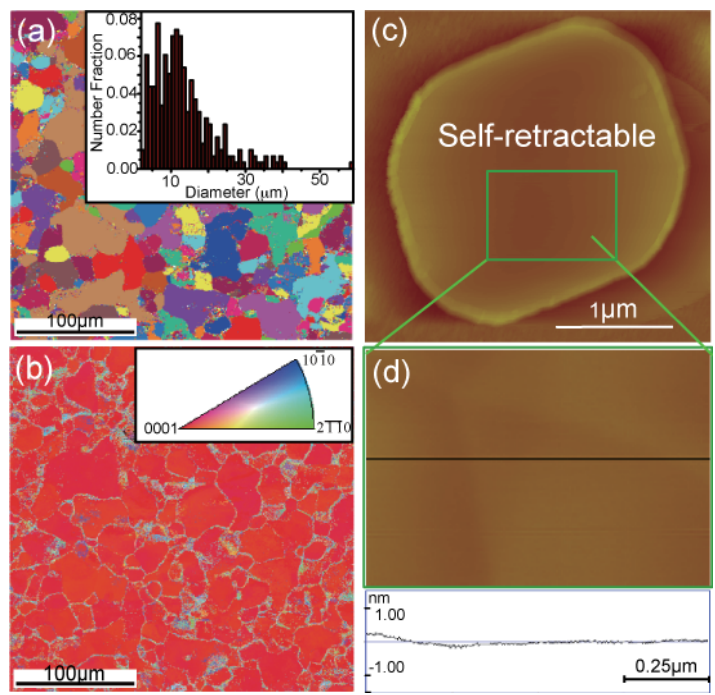

**Fig. 3** (a) Color map of graphite grain orientation from EBSD experiments with different colors corresponding to different grains. The histogram of grain size, where the diameter of each grain is defined as the mean square root of its area, is shown in the inset. The average grain size is ~10$\mu$m. (b) Crystal-direction map (IPF mapping image) from EBSD with the crystal orientation represented by the inverse pole figure in the inset. (c)-(d) STM scan of a typical contact interfaces of a graphite island that exhibits self-retraction, which clearly indicates atomically smooth surface in the self-retraction islands (the surface roughness is below 0.5 nm over a length of 1 μm).

By chance, we obtained a few mesas containing only one incommensurate contact interface. After shearing out the upper flake of such a mesa and rotating it to a lock-in orientation, the obtained sample is an ideal form to determine the ISS for single crystalline graphite. The created locked-in contact area (AB-stacking) is smaller than others in the single crystal grains of the mesa sample (also AB-stacking), which will ensure the unlocking of AB-stacking occurred at the newly created contact interface.

The microtip is then used to shear the top flake of the graphite mesa. From the deformations of the microtip in quasi-static loading and unloading, we can estimate the interlayer shear strength for single crystal graphite. The position of the tip and the loading velocity can be precisely controlled with an accuracy of the micromanipulator (Kleindiek) up to 5 nm. Fig. 4(a) and Fig. 4(b) show such a mesa that is self-retractable. Comparing the shapes of the probe in loading (Fig. 4a) and unloading (Fig. 4b), no detectable deformation is observed with the optical microscope resolution. Shearing a flake from a square graphite mesa of edge length $L$ to a distance $x$ creates new surfaces of total area $2Lx$ (Fig. 1 (d)) and thus an excess surface free energy of $U = 2\gamma Lx$, where $\gamma$ is the graphite basal plane surface energy which is estimated to be about 0.1-0.15 J/m$^2$ [27-28]. The corresponding self-retraction force is thus $F_{\text{retract}} = |-dU/dx| = 2\gamma L$. The friction resistance force is $F_{\text{f}} = \tau_s L(L-x)$. Since the flake self-retracts, we can conclude that this force overcomes the friction at the interface. So we obtain an upper bound estimate of the areal friction stress as $\tau_s^{\text{upper}} = 2\gamma/(L-x_{\text{max}})$, where $x_{\text{max}}$ is the maximum sheared distance in our experiments, typically ~5 μm for a 10 μm mesa. This analysis yields the upper bound estimate $\tau_s^{\text{upper}} \approx$ 0.04-0.06 MPa.

For a lock-in state, Fig. 4(c) is the frame just before the top flake begins to move (selected from an *in-situ* Movie). After unloading, the microtip remains bent (Fig. 4(d)), indicating a plastic deformation has occurred while attempting to shear the top flake. The critical shearing resistance force in the lock-in state is equal to the force exerted to the microtip from the top flake. That latter force can be estimated from the deformation of the microtip in Fig. 4 (c) by finite element modeling (FEM).

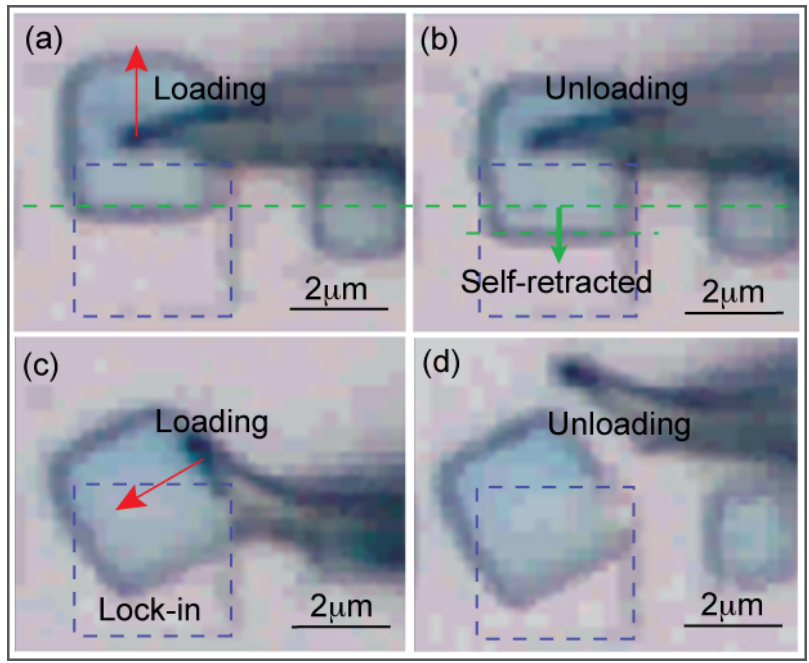

**Fig. 4** (a)-(b) Two image frames taken from in-situ Movie. Shearing a self-retractable graphite flake using the microtip of a micromanipulator (Kleindiek) within an optical microscope (OM, HiRox KH-3000). No obvious deformation of the microtip was observed within the resolution of the optical microscope. (c)-(d) two selected image frames from an in-situ Movies. Shearing a top flake in a lock-in state. Image (c) records the moment just before the top flake begins to move. After unloading, a plastic deformation of the microtip can be seen.

## 3 Finite Element Analysis and Results

In our FEM model (Fig. 5(a)), the top flake is considered as a rigid body, since the deformation of the top flake can be neglected in comparison to the deformation of the microtip as observed in experiments. A classic elastic–linear plastic model with isotropic hardening (MISO options in ANSYS) was used for the mechanical analysis of the tungsten microtip. The elastic modulus of the microtip is 201 GPa, measured from the uni-axial tensile stress-strain curve (Fig. 5 (b)) of a tungsten thread (raw material to fabricate the microtip in our experiments) with a diameter 0.30 mm (WDW3020, Changchun Branch, a strain rate of about $2 \times 10^{-4} \text{s}^{-1}$). The Poisson ratio is selected as 0.284 [29]. Displacement controlled loading is implemented to the microtip in the far end. Different displacement values result in different shapes of neutral surface of the microtip (theoretical shapes) as shown in Fig. 5(c). The red-dotted line is the neutral surface measured from experiments (Fig. 4(c)). Through a comparison, the best fitting yields the critic resultant force exerted to the microtip as $F_c \sim 0.65$ mN. The resultant force is perpendicular to the loading edge of the top flake because the contact between the microtip and the top flake is considered as frictionless in our FEM models.

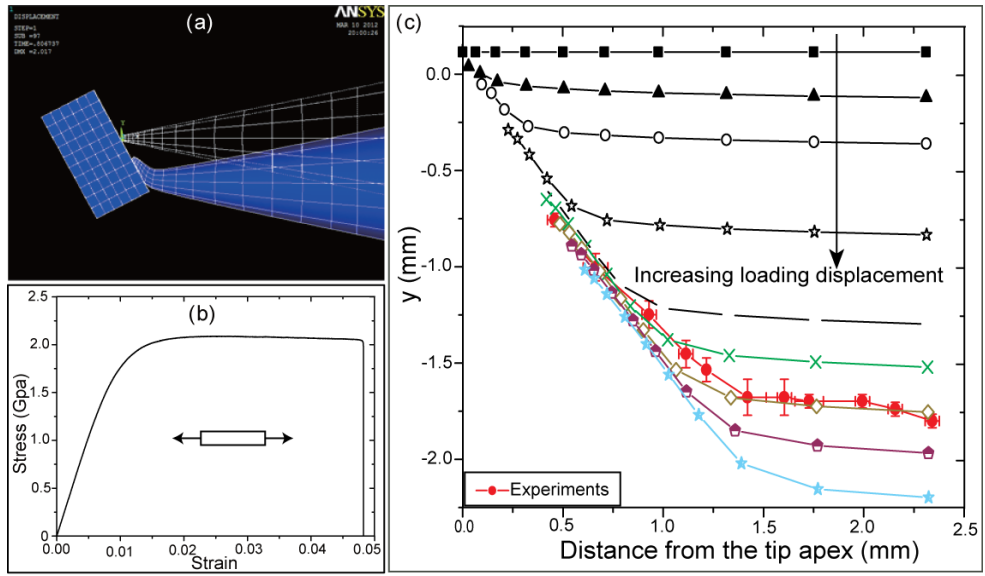

**Fig. 5** (a) FEM model simulating the loading process in experiments (using ANSYS code, see text for details). (b) Stress–strain curve of a tungsten thread with diameter 0.30 mm subject to a uni-axial quasi-static tension ( WDW3020, Changchun Branch, strain rate ~ $2 \times 10^{-4} s^{-1}$). (c) The comparison between the shapes of neutral surface of the microtip at different displacements in FEM models (theretical shapes) and the experimental results (measured from Fig. 4(c) and denoted as red-dot line). The optimal fitting yields the critic resultant force exerted to the microtip.

With the determined critical force and the contact area between the graphite/$SiO_2$ flake and the graphite platform measured in our experiment $S$=4.5 $\mu m^2$, we can estimate the ISS of the AB-stacked graphite layers as $\tau_{s,lock-in} = 0.14$GPa. This value is about two orders of magnitude higher than those measured in a macroscale shear experiments (0.25-2.5 MPa). We believe that the much smaller ISS measured in the previous experiments can be attributed to the existence of many incommensurate contacts in their samples, which could significantly reduce the shear strength because of the superlubricity.

In summary, a novel experimental method is presented to directly measure and estimate the interlayer shear strength for single crystal graphite as $\tau_s \sim 0.14$ GPa. This value is about two orders of magnitude higher than the measured values in previous macro-scale shear experiments (0.25-2.5 MPa). We attributed the drastic difference to the presence of stacking faults (incommensurate contact) in the samples used in the macro-scale experiments. Our results can serve as a benchmark for the shear strength of single crystal graphite.

**Acknowledgment**

Q.S.Z. acknowledges the financial support from NSFC through Grant No. 10832005 and the

National Basic Research Program of China (Grant No. 2007CB936803).